\documentclass[sigconf]{acmart}
\setcopyright{none}
\renewcommand\footnotetextcopyrightpermission[1]{}

\usepackage{booktabs}

\begin{document}

\title{Was It Never Collected, or Rewritten Away?\\
       A Commit-Provenance Dataset Separating Ingestion Gaps from
       Upstream History Edits across the World of Code}

\author{Audris Mockus}
\affiliation{%
  \institution{University of Tennessee, Knoxville}
  \city{Knoxville}\state{TN}\country{USA}}
\email{audris@utk.edu}

\begin{abstract}
Any global mirror of open-source version control is incomplete, but the reasons
a commit is missing are not interchangeable: a project may have force-pushed it
away (so it no longer exists upstream), or the mirror may never have ingested it
(a true collection gap). We release a commit-provenance dataset that separates
these two cases at scale by comparing two views of the same commit graph: the
GHArchive event stream as a historical witness of what GitHub advertised at push
time, and the World of Code (WoC) \texttt{V2604} object database, an accumulated
union of periodic fetches that never deletes a commit once collected, as what the
corpus has gathered to date. Walking each reference's \texttt{PushEvent} chain reconstructs force-push
events structurally (a \texttt{before} that is not the prior \texttt{head} breaks
the fast-forward chain; no recorded flag is needed and none exists across eras),
and joining every advertised commit against WoC membership yields a three-way
label. Over $1{,}118{,}116{,}350$ advertised commits, $53.35\%$ are
\emph{present} in WoC, $6.47\%$ are \emph{rewritten} (absent now and orphaned by
a later force-push on the same reference, an upstream history edit and a correct
absence), and $40.18\%$ are \emph{never-ingested} (absent and never force-push
orphaned, the candidate collection gap). About one missing-commit case in fifteen
is therefore a rewrite the project itself erased rather than a mirror gap. We
also release the force-push witness ($166{,}710{,}831$ events over $19{,}926{,}250$
repositories) and a per-project rollup ($78{,}125{,}788$ repositories;
force-pushing observed in $25.47\%$, a rewritten commit in $12.85\%$). We treat
the $40\%$ never-ingested share as an upper bound on the collection gap and state
the five reasons it overcounts. The dataset makes mirror-completeness reporting
honest, flags rewritten commits as same-patch duplicate sources for contribution
counting, and corrects a $10.82\%$ corpus-wide undercount that history editing
imposes on commit-based productivity, for which we release a per-project correction
factor. All artifacts are released as a self-contained,
independently hosted replication package keyed to the WoC \texttt{V2604}
collection.
\end{abstract}


\maketitle

\section{Introduction}
\label{sec:intro}
Every corpus that mirrors open-source version control at scale is incomplete, and
every such corpus reports a count of commits it expected but does not hold. That
count conflates two phenomena that have opposite implications for an analyst. A
commit can be missing because the upstream project \emph{rewrote its history},
force-pushing, rebasing, squashing, or running a history filter, so the commit no
longer exists anywhere and its absence is correct. Or a commit can be missing
because the mirror \emph{never ingested it}, a genuine collection gap that bounds
what any study built on the corpus can observe. A raw missing-commit number
cannot distinguish the two, so it overstates the collection gap by however much
upstream rewriting contributes, and it silently hides rewritten commits that, when
present elsewhere as re-created SHAs, inflate contribution counts.

\begin{figure}[t]
\centering
\includegraphics[width=\linewidth]{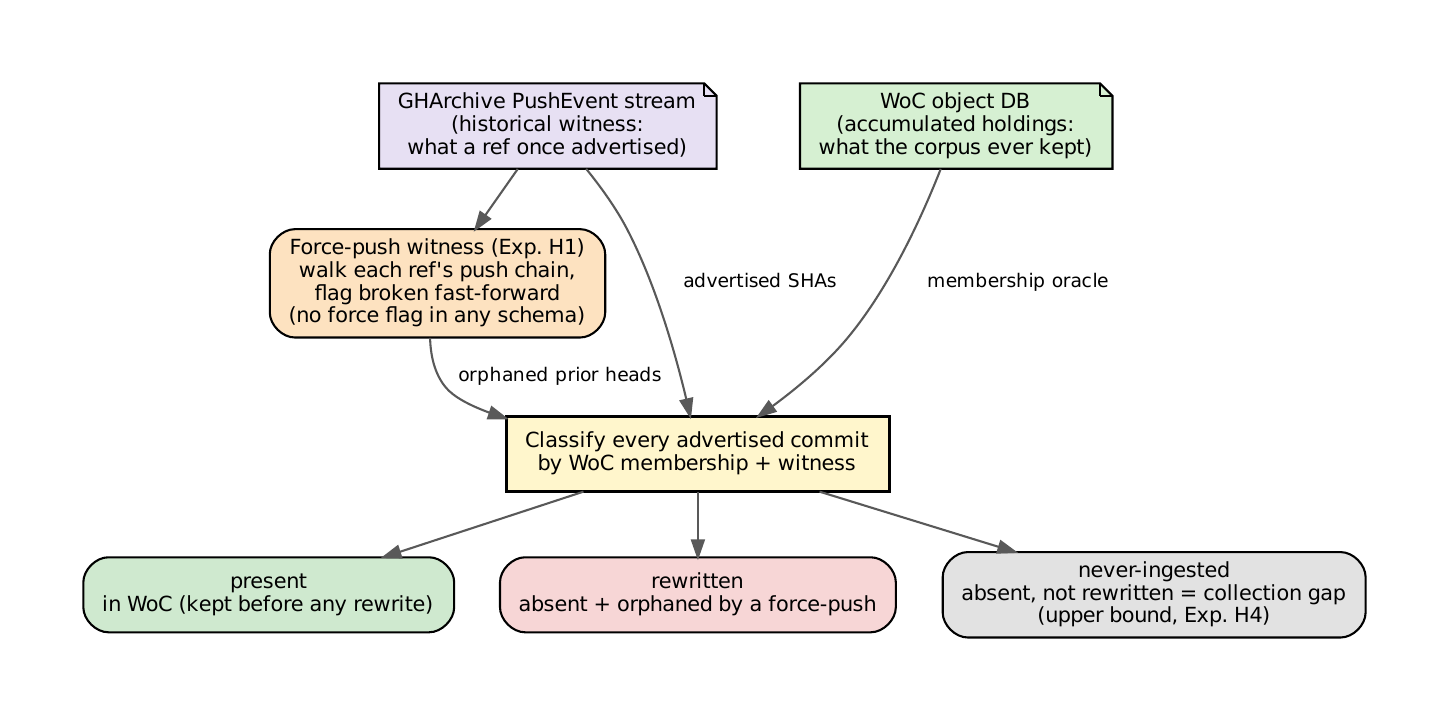}
\caption{Two independent views of the same commit graph classify every advertised
commit. The GHArchive event stream supplies the advertised SHAs and a force-push
witness; WoC supplies the membership oracle. The resulting label is \emph{present},
\emph{rewritten}, or \emph{never-ingested}.}
\label{fig:provenance}
\end{figure}

The two phenomena can be separated because two independent views of the same
commit graph exist. GHArchive~\cite{gharchive} is an append-only log of GitHub's
public event stream; each \texttt{PushEvent} records what a reference advertised
at the moment of a push, including the commit SHAs involved. World of
Code~\cite{ma2019woc,ma2021world} is an accumulated object database: it grows by
periodic fetches of public repositories, never deletes an object once collected,
and so holds the union of everything it has ever observed, with the recency of
that observation varying from one repository to the next. The historical witness
says what was once advertised; the accumulated database says what the corpus has
ever gathered and kept. Comparing them, commit by commit, classifies every
absence (Figure~\ref{fig:provenance}). Because WoC retains a commit even after the upstream project rewrites it
away, a rewritten commit that WoC fetched before the rewrite stays \emph{present},
which keeps the present class large and the inferred gap conservative.

This paper releases that classification for the WoC \texttt{V2604} collection,
together with the intermediate force-push witness that drives it, and documents
both its construction and its empirical results as an experiment log
(Exps.~H1--H5). Our contributions are the released artifacts and the findings
about them:
\begin{itemize}
\item \textbf{A force-push witness} reconstructed from the GHArchive
  \texttt{PushEvent} stream alone, with no reliance on a recorded force flag
  (which is absent in every schema era), by walking each reference's push chain
  and flagging a broken fast-forward (Exp.~H1).
\item \textbf{A three-way commit-provenance map} labeling every advertised commit
  \emph{present} / \emph{rewritten} / \emph{never-ingested} against WoC membership,
  the central data contribution (Exp.~H2).
\item \textbf{Project-level and stream-level incidence} of force-pushing and of
  each provenance class, quantifying how common history rewriting is and how much
  commit volume it actually erases (Exp.~H3).
\item \textbf{A completeness correction and a caveat budget}: the never-ingested
  share is an upper bound on the collection gap, and we account for the three
  reasons it overcounts so downstream users can read it correctly (Exp.~H4).
\item \textbf{A productivity correction} derived from the rewritten class: a
  commit-count productivity measure read off the current visible history
  undercounts real output by $10.82\%$ corpus-wide, and we release a per-project
  correction factor that restores it (Exp.~H5).
\end{itemize}

The dataset is a drop-in join key for any WoC-scale study: per-project counts join
the project summary, and per-commit labels join the commit tables by SHA. Because
it is built from an external historical witness rather than the mirror's own
holdings, it measures the mirror against something the mirror cannot see by
introspection.

\section{Related Work}
\label{sec:related}
The corpus we measure is World of Code~\cite{ma2019woc,ma2021world}, a mirror of
the universe of public version-control data, organized as commit, tree, blob, and
relation maps. Like every mirror it is incomplete, and the perils of treating any
GitHub-derived corpus as complete are well documented:
Kalliamvakou et al.~\cite{kalliamvakou2014promises} show that much GitHub activity
is invisible or misleading in the recorded metadata, and that declared structure
need not match observed history. GHTorrent~\cite{gousios2012ghtorrent,
gousios2013ghtorrent} mirrors GitHub's event API and is itself subject to the same
gap-versus-edit ambiguity for any commit it lacks. Our contribution is to resolve
that ambiguity for a specific corpus by triangulating against an independent
historical witness.

That witness is GHArchive~\cite{gharchive}, the public archive of GitHub's event
timeline. Prior mining work uses GHArchive mostly as a source of activity events
(stars, forks, issues, pushes) treated as facts about the present. We instead use
the \texttt{PushEvent} stream as a temporal record of what each reference once
advertised, and exploit the difference between that record and the current object
database. Reconstructing force-pushes from the push chain, rather than from a
recorded flag, is necessary because no force indicator is present across the
schema eras we span.

The closest sibling construction is the shared-commit deforking map for World of
Code~\cite{forks20} and its GHArchive-validated successor~\cite{deforkingshowcase}:
both reason about commit sharing across repositories, and both confront the same
completeness question from the project-graph side. That companion paper validates
a commit-based deforking map against GitHub's declared fork graph reconstructed
from the GHArchive \texttt{ForkEvent} stream, the fork-graph view of the same
ingestion-completeness question this paper addresses at the commit level. Where
deforking asks which repositories are the same project, this dataset asks which
advertised commits the mirror holds and why the rest are absent. The rewritten class is also relevant to repository deduplication
work~\cite{spinellis2020dedup}: a rewritten commit re-created with a new SHA after
a rebase is a same-patch duplicate, the commit-level analogue of the
repository-level duplication those resources remove.


\section{Construction}
\label{sec:construction}
The dataset is built from two inputs: the GHArchive \texttt{PushEvent} stream
(the historical witness) and the WoC \texttt{V2604} commit tables, the
accumulated union of everything the corpus has fetched and retained up to that
watermark. Construction runs entirely on the analytics cluster with the corpus
conventions (\texttt{;}-separated, gzip-compressed, \texttt{LC\_ALL=C} sorted),
in four stages.

\paragraph{Extraction.} For every month of the GHArchive mirror, a pre-filter
selects \texttt{PushEvent} records and emits one row per push,
\texttt{created\_at;repo;ref;before;head;forced;ncommits;commit\_shas}, carrying
the repository as its raw GitHub \texttt{owner/repo} name. Normalization to the
WoC key (lowercase, \texttt{/}$\rightarrow$\texttt{\_}) is deferred to the
rollup, where it lets the per-project counts join existing project signals.
The \texttt{PushEvent} schema shifts across eras, and the extractor handles each
transparently: \texttt{before} and \texttt{head} are present from 2015 onward; the
per-commit \texttt{commits[]} array is present 2015--2025 and stripped in 2026
(recent months carry only reference and head/before identifiers); the 2011--2014
timeline carries a \texttt{shas[]} array and a \texttt{head} but no \texttt{before}.
A recorded \texttt{forced} flag is absent in every era, so force-pushing is
recovered structurally rather than read from the event. The advertised
\texttt{head} is present in all eras, so the membership test that drives the
provenance label works corpus-wide; the per-commit array, where present, only
enlarges the set of advertised SHAs.

\paragraph{Force-push reconstruction.} Pushes are resharded by repository hash and,
within each (repository, reference), ordered by time. A normal push extends the
reference, so its \texttt{before} equals the previous push's \texttt{head} (a
fast-forward chain). A force-push breaks that chain: the new \texttt{before} is not
the prior \texttt{head}. Each break is recorded in \texttt{forcepush.gz} as
\texttt{repo;ref;time;prior\_head;new\_before}, and the set of orphaned
\texttt{prior\_head} SHAs is carried forward as the witnesses of upstream rewrites.
For example, three pushes on one reference with \texttt{before}/\texttt{head}
pairs $(\varnothing, A)$, $(A, B)$, $(A, C)$ read as follows. The second extends
the first, since its \texttt{before} equals the standing head $A$, a
fast-forward. The third declares \texttt{before}${=}A$ while the reference
already stood at $B$, so the chain breaks: $B$ is orphaned as a rewrite witness,
and the commits unique to $B$ were advertised and then erased from the
reference's history.

\paragraph{Provenance join.} The union of all advertised SHAs (each push's
\texttt{head}, plus the \texttt{commits[]} fan-out where present) is resharded by
SHA first byte to match the WoC commit tables and joined against WoC membership.
Each advertised SHA receives one of three labels, written to
\texttt{prov.b\{0..127\}.gz} as \texttt{sha;repo;class}:
\emph{present} if the SHA is in WoC; \emph{rewritten} if it is absent and appears
among the orphaned \texttt{prior\_head} set (an upstream history edit covered it);
\emph{never} if it is absent and was never orphaned by a force-push.

\paragraph{Rollup.} Per-commit labels are aggregated per repository into
\texttt{P2compaction.gz}, \texttt{repo;nForcePush;nRewritten;nNever;nPresent},
which drops in beside the fork and star popularity signals in the project summary.
A post-build quality sweep tests every released file (\texttt{pigz -t} on the
witness, rollup, and all 128 provenance shards: zero corrupt) and confirms the
rollup row count, following the corpus rule that a present file is not assumed
complete until verified.

\section{Force-Push Witness (Exp.~H1)}
\label{sec:h1}
Walking the push chains yields \texttt{forcepush.gz}: $166{,}710{,}831$
force-push events spread over $19{,}926{,}250$ repositories and $25{,}288{,}938$
distinct (repository, reference) pairs. A reference that is ever force-pushed is
force-pushed about $6.6$ times on average, so force-pushing recurs on the
references where it happens rather than being a one-time event. Reconstructing the
witness from the chain structure, with no reliance on a recorded flag, is what
makes the dataset span the full timeline: the same procedure labels a 2012 push
and a 2026 push, even though only one of them would carry per-commit SHAs and
neither carries a force indicator.

One benign source of chain breaks is the witness itself rather than the project.
GHArchive has its own missing hours and months, and a gap in the event stream can
leave a reference's recorded \texttt{before} pointing at a head the archive never
logged, mimicking a force-push. Such a spurious break orphans a commit that was
in fact a normal tip, which an accumulating mirror almost always already holds, so
at the provenance join the orphan matches WoC membership and is labeled
\emph{present} rather than \emph{rewritten}. Spurious breaks therefore do not
inflate the rewritten share or the headline gap; the only residual risk is the
rare orphan that WoC never collected, which would count as rewritten rather than
never, and that case is bounded by the archive's small gap rate.

\section{Three-Way Commit Provenance (Exp.~H2)}
\label{sec:h2}
The central result classifies every commit any \texttt{PushEvent} ever advertised
against WoC membership. Table~\ref{tab:provenance} reports the split over
$1{,}118{,}116{,}350$ advertised commits.

\begin{table}[t]
\centering\small
\caption{Three-way provenance of every advertised commit, classified against WoC
\texttt{V2604} membership. \emph{Rewritten} and \emph{never} are the two reasons
an advertised commit is absent; separating them is the contribution.}
\label{tab:provenance}
\begin{tabular}{@{}lrr@{}}
\toprule
class & commits & share \\
\midrule
present (in WoC)                 & $596{,}497{,}408$   & $53.35\%$ \\
rewritten (force-push orphan)    & $72{,}357{,}948$    & $6.47\%$  \\
never-ingested (candidate gap)   & $449{,}260{,}994$   & $40.18\%$ \\
\midrule
advertised total                 & $1{,}118{,}116{,}350$ & $100\%$ \\
\bottomrule
\end{tabular}
\end{table}

\begin{figure}[t]
\centering
\includegraphics[width=0.92\linewidth]{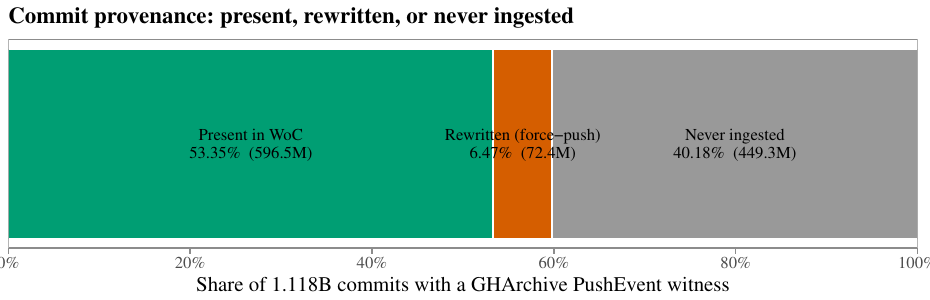}
\caption{Provenance of every advertised commit, as a share of the
$1{,}118{,}116{,}350$ commits carrying a GHArchive \texttt{PushEvent} witness.
Present commits are in WoC; the absent remainder splits into \emph{rewritten}
(force-push orphans) and \emph{never-ingested} (the candidate ingestion gap).}
\label{fig:provenance-split}
\end{figure}

A little over half of advertised commits are present (Figure~\ref{fig:provenance-split}). Of the absent remainder,
the dataset separates the two reasons an analyst cares about: $72{,}357{,}948$
commits ($6.47\%$ of all advertised) were rewritten away upstream, and
$449{,}260{,}994$ ($40.18\%$) are candidate collection gap. Roughly one absent
commit in fifteen is therefore a rewrite the project itself erased, not a mirror
failure. Without the GHArchive witness these two would be a single
``missing'' bucket, and the collection gap would read as $46.65\%$ rather than its
true upper bound of $40.18\%$.

\section{Project- and Stream-Level Incidence (Exp.~H3)}
\label{sec:h3}
The per-project rollup covers $78{,}125{,}788$ repositories, one row per
repository ever seen in a \texttt{PushEvent}. Table~\ref{tab:incidence} reports
how many carry each phenomenon. Force-pushing is common at the project level: a
quarter of all repositories ever observed pushing have force-pushed at least once.
But it erases only a small slice of commit \emph{volume} ($6.47\%$ from
Table~\ref{tab:provenance}), so most force-pushes rewrite a short tip rather than
deep history.

\begin{table}[t]
\centering\small
\caption{Per-project incidence over $78{,}125{,}788$ repositories. Counts are
repositories with at least one event of each type.}
\label{tab:incidence}
\begin{tabular}{@{}lrr@{}}
\toprule
repositories with\ldots & count & share \\
\midrule
a force-push                  & $19{,}897{,}479$ & $25.47\%$ \\
any rewritten commit          & $10{,}036{,}933$ & $12.85\%$ \\
any never-ingested commit     & $46{,}226{,}356$ & $59.17\%$ \\
any present commit            & $31{,}827{,}928$ & $40.74\%$ \\
\bottomrule
\end{tabular}
\end{table}

The force-push project count from the rollup ($19{,}897{,}479$) sits $28{,}771$
below the distinct-repository count in the witness ($19{,}926{,}250$). The gap is
repositories whose force-push rows carry no advertised SHA that survived into the
provenance keying, a rounding-level discrepancy rather than a defect.

\section{Completeness Correction and Caveat Budget (Exp.~H4)}
\label{sec:h4}
The dataset's practical use is to make completeness reporting honest, so the
$40.18\%$ never-ingested figure must be read correctly. It is an \emph{upper
bound} on the true collection gap, for five reasons that all push it high rather
than low.

\begin{enumerate}
\item \textbf{Watermark.} Membership is tested against the \texttt{V2604}
  snapshot. Commits pushed after the watermark are correctly labeled never with
  respect to that snapshot but are not a collection gap; a later watermark would
  reclassify them.
\item \textbf{Pre-2015 blind spot.} The 2011--2014 timeline carries no
  \texttt{before}, so the force-push chain cannot be walked there. A commit
  rewritten away in that era cannot be reclassified as rewritten and falls into
  never. The never bucket therefore absorbs old-timeline rewrite orphans that, on
  a complete witness, would be rewritten.
\item \textbf{Resolved-table oracle.} Membership is tested against the resolved
  commit table rather than the fuller raw object-presence oracle, a choice made to
  keep construction on the analytics cluster. A commit present in the raw object
  database but absent from the resolved table reads as absent here, making never
  conservative-high.
\item \textbf{Retrieval-cadence lag.} WoC fetches each repository periodically, so
  per-repository observation recency varies. A commit pushed not long before the
  watermark, on a repository WoC has not re-fetched since, is live upstream yet not
  yet collected, so it reads as never though a later fetch would gather it. This is
  a timing artifact, not a permanent gap, and it inflates never.
\item \textbf{Non-force-push history loss.} History can disappear with no
  \texttt{before}/\texttt{head} fingerprint to witness it: a branch or whole
  repository deleted, or a repository made private, removes commits without leaving
  a broken fast-forward in the push chain. Such orphans carry no rewrite evidence
  and fall into never, so never absorbs upstream history edits that left no
  structural trace.
\end{enumerate}

All five move probability mass from present or rewritten into never, so the true
collection gap is at most $40.18\%$ and the rewrite share is at least $6.47\%$.
For an analyst this is the actionable correction: the headline missing-commit
number overstates what the mirror could ever have collected, and a measurable
slice of the apparent gap is upstream history that no longer exists to be
collected. A second use follows from the rewritten class directly: a rewritten
commit re-created with a fresh SHA after a rebase or squash is a same-patch
duplicate, so contribution and provenance analyses can use the rewritten labels to
avoid counting a rebased commit and its replacement as two distinct contributions.

\section{Editing Impact on Measured Productivity (Exp.~H5)}
\label{sec:h5}
The rewritten class corrects a bias in commit-based productivity, a use that
follows directly from the provenance labels. A productivity measure read off the
current visible history counts only the commits a reference still reaches, the
\emph{present} commits; commits a later force-push orphaned are invisible to it, so
per-project commit output is undercounted by exactly the rewritten share of a
project's true output. Aggregated over the corpus, a naive current-history count
sees the $596{,}497{,}408$ present commits where the reachable-plus-rewritten total
is $668{,}855{,}356$ (Table~\ref{tab:provenance}), an editing-induced undercount of
$10.82\%$: roughly one commit in nine of all real output is masked by history
editing alone. This is separate from the collection gap, which concerns commits the
mirror never held rather than output a project produced and then erased. Excluding
the $440{,}594$ force-pushes that straddle a GHArchive outage
(Section~\ref{sec:h1}), which a conservative reading drops as archive artifacts
rather than genuine rewrites, lowers the figure only to $10.76\%$; the correction
is thus robust to how archive gaps are treated.

The loss is concentrated rather than diffuse. Restricting to the $40{,}584{,}004$
repositories with a defined edit rate (at least one present or rewritten commit;
repositories whose entire advertised history is never-ingested have no visible
output to correct and are excluded), three-quarters saw no editing at all, but
among those that did the visible count understates true output by $37.95\%$. More
pointedly, $8{,}689{,}485$ of them ($21.41\%$) have an \emph{empty} visible history
despite real prior output: every commit they ever advertised was later force-pushed
away, so a current-history productivity read is exactly zero for a repository that
did produce commits. Only the rewritten witness recovers them.

We release the per-project correction as \texttt{P2prodCorr.gz}
(\texttt{project;visible;true\_gap;true\_raw;undercount\_gap;corr\_factor\_gap}):
multiplying any per-project commit-productivity metric by \texttt{corr\_factor\_gap}
(or adding back \texttt{true\_gap$-$visible}) restores the commits history editing
removed, with \texttt{corr\_factor\_gap}${=}\infty$ marking the empty-history
repositories above. The correction recovers commit \emph{count}, not author
attribution: an orphaned commit is not re-attributed to its developer here, which
would require joining the orphaned SHAs back to their authors, a follow-on left to
the per-commit provenance shards.

\section{Limitations}
\label{sec:limitations}
Beyond the caveat budget of Exp.~H4, two scope limits bound the dataset. First,
GHArchive only records \emph{public} activity at the moment it occurred: a
repository made public after a history rewrite shows only its post-rewrite state,
so the rewrite is invisible to the witness. Second, the witness covers commits an
event stream advertised; commits that reached WoC through channels GitHub never
emitted a \texttt{PushEvent} for (other forges, direct ingestion) are present in
WoC but never appear in the advertised universe, so they are outside the
classification rather than mislabeled. Both limits are conservative for the
headline claim: they can only hide rewrites or shrink the advertised universe,
not inflate the rewritten share.

\section{Availability}
\label{sec:availability}
All artifacts are released as a single self-contained bundle; no World of Code
account is required to obtain or use them (artifact schemas, keys, and row counts
are summarized in Table~\ref{tab:artifacts}): the per-project rollup
\texttt{P2compaction.gz}, the force-push witness \texttt{forcepush.gz}, the
per-commit provenance shards \texttt{prov.b\{0..127\}.gz}, and the per-project
productivity correction \texttt{P2prodCorr.gz} (Exp.~H5), each
\texttt{;}-separated, gzip-compressed, and \texttt{LC\_ALL=C} sorted, together
with a replication package that regenerates them from the WoC commit tables and
the GHArchive \texttt{PushEvent} mirror. Because the witness and provenance files
are multi-gigabyte, the data artifacts are hosted as a Hugging Face dataset (which
scales past the per-file limits of a code host and mints a citable DataCite DOI),
with the replication code mirrored on GitHub and cross-linked from the dataset
card. We will also offer the identical bundle to the World of Code maintainers,
so that existing WoC users could obtain it through the channels they already use,
should the maintainers choose to adopt it. The data is released under
\textsc{CC-BY-4.0} and the replication code under the \textsc{MIT} license.
\emph{(Hugging Face dataset: \texttt{TODO/woc-history-rewrite-v2604}; DOI to be
minted on camera-ready.)}

\begin{table*}[t]
\centering\small
\caption{Released artifacts (WoC \texttt{V2604}), classified against the
GHArchive \texttt{PushEvent} stream. All files are \texttt{;}-separated,
gzip-compressed, and \texttt{LC\_ALL=C} sorted on the first column; row counts
are measured from the released files. The \emph{key} column names each artifact's
join key.}
\label{tab:artifacts}
\begin{tabular}{@{}llllr@{}}
\toprule
artifact & schema (columns) & key & sharding & rows \\
\midrule
\texttt{P2compaction.gz} & \texttt{repo;nForcePush;nRewritten;nNever;nPresent} & \texttt{repo} & single file & $78{,}125{,}788$ \\
\texttt{forcepush.gz}    & \texttt{repo;ref;time;prior\_head;new\_before}      & \texttt{repo} & single file & $166{,}710{,}831$ \\
\texttt{prov.b\{0..127\}.gz} & \texttt{sha;repo;class} (present/rewritten/never) & \texttt{sha} & 128 SHA-byte shards & $1{,}118{,}116{,}350$ \\
\texttt{P2prodCorr.gz} & \texttt{project;visible;true\_gap;true\_raw;undercount\_gap;corr\_factor\_gap} & \texttt{project} & single file & $40{,}584{,}004$ \\
\bottomrule
\end{tabular}
\end{table*}

\section{Conclusion}
\label{sec:conclusion}
A mirror's missing-commit count answers the wrong question, because it merges
commits the mirror failed to collect with commits the upstream project erased. By
comparing the GHArchive push witness against the World of Code object database, we
label every advertised commit present, rewritten, or never-ingested, and find that
roughly one missing case in fifteen is an upstream history edit rather than a
collection gap. The released provenance map, force-push witness, per-project
rollup, and productivity correction let any WoC-scale study report completeness
honestly, treat rewritten commits as the same-patch duplicates they are, and
recover the $10.82\%$ of commit output that history editing hides from a
current-history productivity count, turning an external event stream into a
measurement the mirror cannot make of itself.

\bibliographystyle{ACM-Reference-Format}
\bibliography{refs}

\end{document}